\documentclass[reprint,bibnotes,aps,prb,floatfix,superscriptaddress,longbibliography]{revtex4-1}
\usepackage{graphicx}
\usepackage{amsmath}
\usepackage{amsxtra}
\usepackage{amssymb}
\usepackage{dcolumn}
\usepackage{float}
\usepackage{bm}
\usepackage[breaklinks=true,colorlinks,citecolor=blue,linkcolor=blue,urlcolor=blue]{hyperref}

\def\nn{\nonumber}
\def\be{\begin{equation}}
\def\ee{\end{equation}}
\def\bea{\begin{eqnarray}}
\def\eea{\end{eqnarray}}
\def \Pis{\Pi_{\sigma_z\sigma_z}}
\def \Pir{\Pi_{\rho \rho}}
\def \hatz{\hat{\bf{z}}}
\def \hatx{\hat{\bf{x}}}

\begin{document}

\title{Dynamic current-current susceptibility in 3D Dirac and Weyl semimetals} 
\author{Anmol Thakur}
\affiliation{Department of Physics, Indian Institute of Technology , Kanpur 208016, India}
\author{Krishanu Sadhukhan}
\affiliation{Department of Physics, Indian Institute of Technology , Kanpur 208016, India}
\author{Amit Agarwal}
\email{amitag@iitk.ac.in}
\affiliation{Department of Physics, Indian Institute of Technology , Kanpur 208016, India}

\date{\today}

\begin{abstract}
We study the linear response of doped three dimensional Dirac and Weyl semimetals to vector potentials, by calculating the wave-vector and frequency dependent current-current response function analytically. The longitudinal part of the dynamic current-current response function is then used to study the plasmon dispersion, and the optical conductivity. The transverse response in the static limit yields the orbital magnetic susceptibility. In a Weyl semimetal, along with the  current-current response function, all these quantities are significantly impacted by the presence of parallel electric and magnetic fields (a finite ${\bf E}\cdot{\bf B}$ term), and can be used to experimentally explore the chiral anomaly.  
\end{abstract}

\maketitle
\section{Introduction}  

Dirac and Weyl semimetals are materials with linearly dispersing bands touching at discrete Dirac/Weyl points \cite{PRL.108.140405,ncomms5898,Sumathi_review,AV_review}. Graphene is one of the most prominent example of a Dirac material in two dimensions (2D). In 3D materials, Dirac points appear due to accidental band crossings, and are robust against gap opening, only if protected by some crystallographic symmetry\cite{PRL.108.140405,ncomms5898}. The presence of time reversal and crystal inversion symmetry forces the Dirac point to be four fold degenerate, with two degenerate pairs of linearly dispersing bands. Breaking of the time reversal (or crystal inversion) symmetry splits the Dirac node into a pair of Weyl nodes of opposite chiralities displaced in momentum (or energy). 3D Weyl fermions have been realized in TaAs\cite{Xu613,Huang_Weyl,Zhang_Weyl,PhysRevX.5.031013,PhysRevX.5.031023,TaAs_Yang}, NbP \cite{CS_NbP_Weyl}, Mo$_x$W$_{1-x}$Te$_2$\cite{MoWTe2_Weyl} and photonic crystals\cite{Lu622}. 3D Dirac semimetals have been realized in Na$_3$Bi \cite{DSM_A3Bi,Liu864,PhysRevB.94.085121}, Cd$_3$As$_2$ \cite{Cd3As2_1,Cd3As2_Natmat,DSM_Cd3As2,PhysRevLett.113.027603,Cd3As2_2} and ZrTe$_5$\cite{ZrTe5_Nphys,PhysRevB.92.075107}.  

A peculiar phenomena related to Weyl semimetals, is the chiral anomaly in crystals: pumping of charges between the nodes of opposite chirality in presence of parallel electric and magnetic fields (finite ${\bf E}\cdot{\bf B}$ term) \cite{NIELSEN1983389}. This non-conservation of the number of particles in a given Weyl node, 
is a direct consequence of the lowest Landau level carrying only right or left movers (depending on the chirality of the Weyl node), as demonstrated explicitly in  Ref.~[\onlinecite{NIELSEN1983389}]. Alternately, it can also be obtained in a semiclassical transport framework as shown in Ref.~[\onlinecite{son_arxiv}], or  from a field theoretic framework of Ref.~[\onlinecite{PhysRevB.96.085201}].
There have been several proposals to  detect the chiral anomaly: in collective density excitations or plasmons \cite{xiao_arxiv_2014,PhysRevB.91.241108}, transport experiments \cite{PhysRevB.88.104412,PhysRevLett.108.046602,burkov,Hosur:2013aa,PhysRevB.93.085426}, optical conductivity \cite{PhysRevB.89.245121}, circular and linear dichroism \cite{kargarian_2015,PhysRevB.91.081106} etc. 

In this paper we study the response of a single node of Dirac and Weyl semimetals to static and dynamic vector fields, by explicitly calculating the current-current response function \cite{principipolini,Stauber,Schliemann}. For each node, we consider a rotationally invariant system in which the current-current correlation function can be expressed as a combination of longitudinal (wave-vector $\parallel$ to the vector field) and transverse (wave-vector $\perp$ to the vector field) response. The longitudinal current-current response function determines the optical conductivity \cite{PhysRevB.87.245131} of the system. It is also related to the density-density response via the current continuity equation and hence determines the dielectric properties and the spectrum of collective density excitations (plasmons) as well\cite{zhang_ijmp_2013,xiao_arxiv_2014,rashi_2015,anmol_2017}. The transverse current-current response function determines the diamagnetic/orbital susceptibility \cite{PhysRevB.81.195431}. 
We present analytical results for the wave-vector and frequency dependent longitudinal as well as transverse current current response function for for a single Dirac node, and then use it to explore the impact of chiral anomaly in Weyl semimetals. In particular the impact of chiral anomaly (a finite ${\bf E}\cdot{\bf B}$ term) can be observed via its impact on the plasmon dispersion, optical conductivity, and the diamagnetic susceptibility. 

The paper is organized as follows: In Sec.~\ref{Sec2}, we set up the calculation of the current-current response function for a single Dirac node. The results of the longitudinal response function are discussed in Sec.~\ref{Sec3}, followed by the results for the transverse case in Sec.~\ref{Sec4}. In Sec.~\ref{Sec5} we study the response of Weyl semimetals in context of the chiral anomaly. Section \ref{new} explores the implications for anisotropic systems,  and we summarize our results in Sec.~\ref{Sec6}.

\section{Current-Current Response function of a single Dirac node}
\label{Sec2}
The effective low energy continuum Hamiltonian to describe a single isotropic massless 3D Dirac (or Weyl) node  
is given by 
\be \label{eqH}
{\mathcal H}= \hbar v_{\rm F} \left(k_x \sigma_x + k_y \sigma_y + k_z \sigma_z \right)~,
\ee
where, $\sigma_i$ are the Pauli matrices denoting real spins, and $v_{\rm F}$ is the Fermi velocity. The response of this system to an electromagnetic vector potential with spatio-temporal variations, ${\bf A(q},\omega)$, is determined by the current-current response function $\Pi_{j_kj_l}(\bf{q},\omega)$. In general $\Pi_{AB}({\bf q},\omega)$ describes the response of the observable $\hat{A}$ coupled to a second observable $\hat{B}$, and  is defined by standard Kubo product \cite{giuliani}~, 
\be\label{response}
\Pi_{AB}(\omega)= -\frac{i}{\hbar S} \lim_{\epsilon \to 0^+} \int_{0}^{\infty} dt\langle[\hat{A}(t),\hat{B}(0)]\rangle e^{i\omega t}e^{-\epsilon t}~, 
\ee
where $S$ denotes the volume of the system. 
For the Hamiltonian given in Eq.~\eqref{eqH}, density operator is given by  $\hat{\rho}_{\bf{q}}=\sum_{\bf{k},\alpha}\hat{\psi}^{\dagger}_{\bf{k-q},\alpha}\hat{\psi}_{\bf{k},\alpha}$  and the corresponding current operator is given by $\hat{j}_{i,q}=v_{\rm F} \sum_{k,\alpha,\beta}\hat{\psi}^{\dagger}_{\bf{k-q},\alpha}{\bf\sigma}_{i,\alpha \beta}\hat{\psi}_{\bf{k},\beta}$. Since the current operator depends on the spin operator, the current-current response function can be expressed in terms of  the spin-spin response function via the relation $\Pi_{j_kj_l}(\bf{q},\omega)=$ $v_{\rm F}^2~\Pi_{\sigma_k \sigma_l}(\bf{q},\omega)$.

The non-interacting spin-spin response function is explicitly given by 
\bea \label{response_func}
\Pi_{\sigma_i \sigma_i}^{(0)}({\bf q},\omega)&=&\frac{1}{S} \lim_{\epsilon \to 0^+}\sum_{k}\sum_{\lambda,\lambda'}\frac{n_{{\bf k},\lambda}-n_{{\bf k+q},\lambda'}}{\hbar\omega + \lambda \varepsilon_{\bf k}-\lambda' \varepsilon_{\bf k+q} +i \epsilon}\nn \\
&\times&|\langle \chi_{\lambda}({\bf k})|\sigma_i|\chi_{\lambda'}({\bf k+q})\rangle|^2~,
\eea
where $\varepsilon_{\bf k}= \hbar v_{\rm F} |{\bf k}|$, $\lambda,\lambda'= 1$ ($-1$) are the band indices for conduction (valence) band and $\chi_{\lambda}({\bf k})$ is the corresponding normalized eigen-spinor.
In general, the spin-spin response function depends on both the magnitude and direction of the wave vector ${\bf q}$. However for systems with  rotational symmetry it can be broken into longitudinal and transverse components and both of them depend only on $q=|{\bf q}|$.  We calculate $\Pi^{(0)}_{\sigma_z \sigma_z}(q,\omega)$ considering $\bf{q}$ along $z$ axis ($\parallel$ to the applied vector field) for the longitudinal part and $\bf{q}$ on the  $x-y$ plane ($\perp$ to the applied field) for the transverse part.

\section{Longitudinal spin-spin response function}
\label{Sec3}
We now discuss the case of non-interacting longitudinal spin-spin response function, {\it i.e.}, $\Pi_{\sigma_z\sigma_z}^{(0)}(q{\hat {\bf{z}}},\omega)$, in both the undoped and the doped scenario.  
Using the current continuity relation, $i\partial_{t}\hat{\rho}_{\bf{q}}=\bf{q}\cdot \hat{j}_{\bf{q}} $, the longitudinal spin-spin response function can be related to the dynamical density-density response function $\Pi_{\rho \rho}(q,\omega)$ via the relation: 
\be \label{den2spin}
\Pi_{\rho \rho}(q \hatz,\omega)=\frac{q}{\hbar\omega^2}\langle [ {\hat j}_{z,q}, \rho_{{-q \hatz}}]\rangle + \frac{q^2 v_{\rm F}^2}{\omega^2}\Pi_{\sigma_z \sigma_z}(q {\bf \hat{z}},\omega)~.
\ee
Here the first term in the r.h.s. of Eq. (\ref{den2spin}) is the anomalous commutator and it arises due to presence of infinite sea of negatively charged electrons in the continuum version of the Dirac Hamiltonian \cite{principipolini}. It turns out to be purely real and is given by 
\be  \label{norm}
\frac{1}{\hbar v_{\rm F}^2 q}\langle [ {\hat j}_{z,q}, \rho_{{-q \hatz}}]\rangle=\frac{q_{\rm max}^2}{6\pi^2\hbar v_{\rm F}}=\frac{\varepsilon_{\rm max}^2}{6\pi^2\hbar^3v_{\rm F}^3}~,
\ee
where $q_{\rm max}$ ($= \varepsilon_{\rm max}/\hbar v_{\rm F}$) is wave vector corresponding to the ultraviolet energy cutoff. For details of the calculation of Eq.~\eqref{norm}, see appendix \ref{AppendixA}. 

Similar to the case of the density-density response function \cite{anmol_2017}, the total non-interacting spin-spin response function can also be expressed as a sum of contributions coming from the undoped part, and an additional doping dependent contribution:  
\be
\Pis^{(0)}(q\hatz,\omega)=\Pis^{(0u)}(q\hatz,\omega)+\Pis^{(0d)}(q\hatz,\omega)~.
\ee
\subsection{Undoped case}
\label{undoped}
For the undoped case,  the Fermi energy lies at the Dirac point and all the contributions are solely  from the inter-band transitions occurring from the full valence band to empty conduction band. Equation~\eqref{den2spin} leads to the following relation between the intrinsic (undoped) parts of the spin and density response functions: 
\be\label{imagden2spin}
\text{Im}\Pi_{\sigma_z\sigma_z}^{(0u)}(q{\hat {\bf{z}}},\omega)=\frac{\omega^2}{ v_{\rm F}^2q^2}\text{Im}\Pi^{(0u)}_{\rho\rho}(q,\omega)~,
\ee
where 
\be \label{eq:Imd}
\text{Im} \Pi^{(0u)}_{\rho\rho}(q,\omega)=-\frac{q^2}{24\pi\hbar v_{\rm F}} \Theta(\omega -v_{\rm F}q)~,
\ee
is the known density-density response function for undoped single Dirac node \cite{xiao_arxiv_2014,zhang_ijmp_2013}. 
In Eq.~\eqref{eq:Imd}, $\Theta(x)$ denotes the step function. Note that the $\omega = v_{\rm F}q$ line also marks the boundary of the intra-band single particle-hole excitations in massless Dirac systems (both doped and undoped).  

The real part of the response function $\text{Re}\Pis^{(0u)}(q \hatz,\omega)$ can be calculated directly using Eq. (\ref{response_func}). Upon integrating Eq. (\ref{response_func}) and after simplification we get
\be \label{reden2spin}
\text{Re}~\Pi_{\sigma_z\sigma_z}^{(0u)}(q{\hat {\bf{z}}},\omega)=-\frac{q_{\rm max}^2}{6\pi^2\hbar v_{\rm F}}+
\frac{\omega^2}{v_{\rm F}^2q^2}\text{Re}\Pi^{(0u)}_{\rho\rho}(q,\omega)~,
\ee
where the real part of the density response is given by\cite{xiao_arxiv_2014,zhang_ijmp_2013},
\be \label{chi_roro}
\text{Re}\Pi_{\rho\rho}^{(0u)}(q,\omega)=-\frac{q^2}{24\pi^2\hbar v_{\rm F}}\log \frac{4v_{\rm F}^2q_{\rm max}^2}{|v_{\rm F}^2q^2-\omega^2|}~.
\ee
Note that Eq.~\eqref{reden2spin} also follows directly from Eqs.~\eqref{den2spin}-\eqref{norm}. 

Before proceeding further, we note that the current current response function for the undoped case corresponds to the polarization bubble diagram of quantum electrodynamics (QED) in 3+1 dimensions. 
In QED - with an infinite energy spectrum- the dimensional regularisation scheme \cite{THOOFT1972189} is generally used and it only gives logarithmic divergence. However in a lattice system, this cannot be implemented and an energy cutoff scheme has to be used. More importantly, condensed matter systems always have a finite bandwidth corresponding to the Bloch bands. In such systems an energy cutoff regularization scheme is also physically relevant, with the interpretation of the cutoff energy scale as the energy bandwidth. 

In quantum electrodynamics (QED) the calculation of the polarization bubble diagram in 3+1 dimensions, with the cutoff regularization scheme, in known to lead to unphysical $q_{\rm max}^2$ divergence. In QED such divergences are generally taken care of by adding a suitable counter term 
in the Lagrangian for cancelling out the divergence. This accounts for the renormalisation of the scale dependent and screened electric charge: $e \to e_r$. See Appendix  \ref{QED} for details. 
A similar interpretation can be made for our work as well. The unphysical quadratically diverging cutoff dependent terms in Eq.~\eqref{reden2spin}, can be taken care of by redefining the effective renormalized scale dependent charge of the Dirac quasiparticles.

\subsection{Doped case}
\begin{figure}[t] 
\begin{center} 
\includegraphics[width=0.9\linewidth]{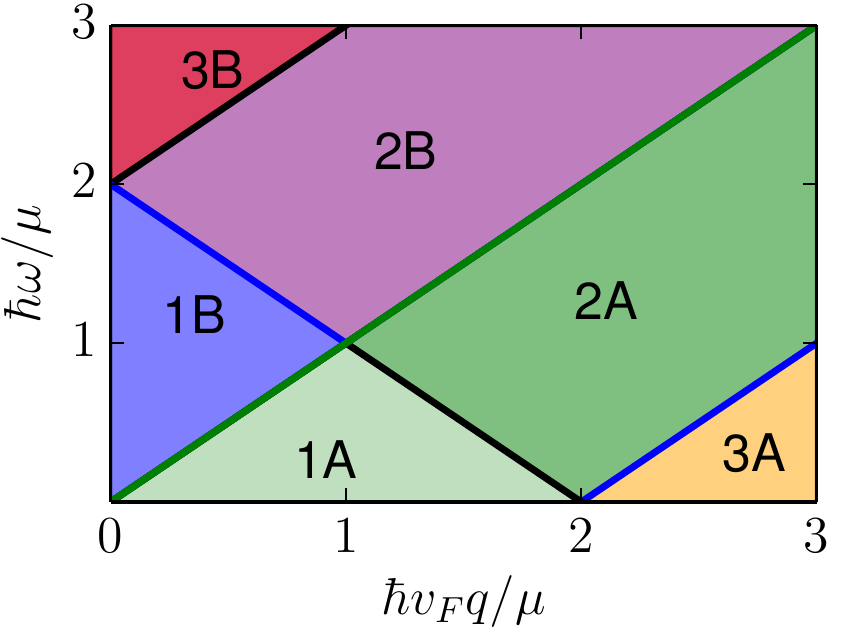} 
\end{center} 
\caption{Different regions in the $\omega -q$ plane used to define the imaginary part of the longitudinal as well as the  transverse current-current response function in a 3D Dirac semimetal. Region 1A and 2A only have intraband particle-hole excitations, while regions 2B has only interband particle-hole excitations. Regions 1B, 3A, and 3B do not have any single particle excitations. 
\label{fig0} } 
\end{figure}

We now proceed to calculate additional contribution to the non-interacting longitudinal spin-spin (or pseudospin-pseudospin) response function $\Pi_{\sigma_z\sigma_z}^{(0d)}(q{\hat {\bf{z}}},\omega)$ which arises only at finite doping (say electron doping).  In this case the  valence band is completely
occupied and conduction band is filled up to the Fermi energy level $\mu =\hbar v_{\rm F} k_{\rm F} > 0$, where $k_{\rm F}$ is the Fermi wave-vector. 
Similar to the case of the undoped part, the doping dependent part of the response function can also be expressed in terms of the corresponding density response function using Eq.~\eqref{den2spin}. For the imaginary part we have 
\bea
\text{Im}~\Pis^{(0d)}(q\hatz,\omega)&=&\frac{\omega^2}{v_{\rm F}^2q^2}\text{Im}~\Pi^{(0d)}_{\rho\rho}(q,\omega)~.
\eea
The imaginary component of the extrinsic part of the density-density response function has already been calculated in Ref.~[\onlinecite{xiao_arxiv_2014,zhang_ijmp_2013}]. Thus the imaginary component of spin-spin response function can 
be expressed as 
\be \label{im13d}
\text{Im}~\Pis^{(0d)}
=-\frac{\omega^2}{8\pi\hbar v_{\rm F}^3q^2}
\begin{cases}
\zeta(q,\tilde{\omega})-\zeta(q,-\tilde{\omega}), & \text{1A} \\
\zeta(q,\tilde{\omega}), & \text{2A} \\
-\frac{q^2}{3}, & \text{1B} \\
-\zeta(-q,-\tilde{\omega}), & \text{2B} \\
0, & \text{3A}, \text{3B}~.\\
\end{cases}~
\ee 
In Eq.~\eqref{im13d}, we have defined the function, 
\bea \label{eq:zeta}
\zeta(q,\omega)&=& \frac{1}{12\hbar^3 v_{\rm F}^3q}\Big[(2\mu+\hbar\omega)^3-3\hbar^2v_{\rm F}^2q^2(2\mu+\hbar\omega)\nn\\
&+&2\hbar^3v_{\rm F}^3q^3\Big]~.
\eea
The various regions in the $\omega-q$ plane, specified by Eq.~\eqref{im13d}  are defined as   
\bea \label{eq:regions}
\text{1A} : & &0<\omega< v_{\rm F}q ~~~~~~~{\rm and}~~~~~~~ 2\mu-\hbar v_{\rm F}q-\hbar\omega>0 ,\nonumber\\
\text{2A} : & &0<\omega< v_{\rm F}q ~~~~~~~{\rm and}~~~~~~~ \pm2\mu\mp\hbar v_{\rm F}q+\hbar\omega>0,\nn\\
\text{3A}: & &0<\omega< v_{\rm F}q ~~~~~~~{\rm and}~~~~~~~ 2\mu-\hbar v_{\rm F}q+\hbar\omega<0, \nn\\
\text{1B} : & &0< v_{\rm F}q<\omega ~~~~~~~{\rm and}~~~~~~~ 2\mu-\hbar v_{\rm F}q-\hbar\omega>0,\nonumber \\
\text{2B} :& & 0< v_{\rm F}q< \omega ~~~~~~~{\rm and}~~~~~~~ \mp2\mu\pm\hbar\omega+\hbar v_{\rm F}q>0,\nn \\
\text{3B} :& & 0< v_{\rm F}q<\omega~~~~~~~{\rm and}~~~~~~~2\mu+\hbar v_{\rm F}q-\hbar\omega<0~. \nn  \\
\eea
These regions are also marked in the $\omega-q$ plane in Fig.~\ref{fig0}. Note that the finite contribution in the 1B region in Eq.~\eqref{im13d}, cancels out the corresponding contribution from the undoped part in Eq.~\eqref{imagden2spin}. Thus in all, regions 1B  and 3A are the only regions without single particle-hole excitations.   

The real part of the doped response function can also be obtained using Eq.~(\ref{den2spin}),  and it is given by  
\be \label{eq:16}
\text{Re}\Pis^{(0d)}(q\hatz,\omega) =\frac{\omega^2}{v_{\rm F}^2q^2}\text{Re} \Pi_{\rho\rho}^{(0d)}(q,\omega)~.
\ee
The real component of extrinsic density-density response function in Eq.~\eqref{eq:16} is given by\cite{xiao_arxiv_2014,zhang_ijmp_2013}, 
\be \label{eq:17}
\text{Re}\Pi_{\rho\rho}^{(0d)}(q,\omega)=-\frac{q^2}{8\pi^2\hbar v_{\rm F}}\left[\mathcal{C}(q,\omega)+\mathcal{D}(q,\omega)\right]~.
\ee 
In Eq.~\eqref{eq:17}, we have defined the following functions: 
\bea \label{eq:CD}
\mathcal{C}(q,\omega)&=&\frac{8\mu^2}{3\hbar^2 v_{\rm F}^2q^2}-\frac{\zeta(q,\omega)H(q,\omega)}{q^2}-\frac{\zeta(-q,\omega)H(-q,\omega)}{q^2}, \nn \\
\mathcal{D}(q,\omega)&=& {\cal C}(q,-\omega)-\frac{8\mu^2}{3\hbar^2v_{\rm F}^2q^2}~, 
\eea
which in turn use $\zeta(q,\omega)$ defined in Eq.~\eqref{eq:zeta} and 
\be
H(q,\omega)=\log \frac{|2\mu+\hbar \omega-\hbar v_{\rm F}q|}{|\hbar v_{\rm F}q-\hbar\omega|}~.
\ee

\subsection{Plasmon dispersion}
The dispersion of the collective density excitations, or plasmon, of an interacting electron gas can be calculated within the random phase approximation. It is given by the zeros of the RPA dielectric function, 
\be\label{eq:18}
\epsilon(q,\omega)=1-v_q\text{Re}\Pir(q\hatz,\omega_{pl})=0~,
\ee
where $v_q=4\pi  e^2/\kappa q^2$ is the 3D Fourier transform of Coulomb potential with $\kappa$ being the surrounding dependent dielectric constant. 
Equation (\ref{eq:18}) can be expressed in terms of spin-spin response function:
\be 
1-v_q\frac{v_{\rm F}^2q^2}{\omega^2}\left[\Pis(q\hatz,\omega)+\frac{q_{\text{max}}^2}{6\pi^2\hbar v_{\rm F}}\right]=0~.
\ee 
Using the calculated spin-spin response function and expanding the expressions in square bracket in Eq.~(\ref{eq:18}) in powers of $q$ upto fourth order, we obtain the plasmon frequency ($\omega_{\rm pl}$) to be  \cite{zhang_ijmp_2013}
\be \label{pl1}
\omega_{\rm pl}=\omega_0\left[1-\frac{\hbar^2v_{\rm F}^2q^2}{8\mu^2}\{1+\gamma(\hbar \omega_0/2\mu)\}\right]~.
\ee 
In Eq.~\eqref{pl1}, we have defined 
\be
 \hbar \omega_0 \equiv \mu \sqrt{\frac{2\alpha_{\rm ee}}{3\pi\kappa^*(\omega_0)}}~,~~~~{\rm and}~~~~ 
~\gamma(x)=\frac{x^2-3/5}{x^2(1-x^2)^2}~,
\ee 
where $\alpha_{\rm ee}=e^2/(\kappa \hbar v_{\rm F})$ is the effective fine structure constant. The effective background dielectric constant of a Dirac node now becomes frequency dependent and it is given by $\kappa^*(\omega)=1+\frac{\alpha_{\rm ee}}{6\pi}\log\left|\frac{4\hbar^2v_{\rm F}^2q_{\text{max}}^2}{\hbar^2\omega^2-4\mu^2}\right|$. This is unlike the case of 2D massless Dirac systems (such as graphene\cite{Wunsch,PhysRevB.75.205418,PhysRevB.91.245407} or 2D surface states of 3D topological insulators \cite{PhysRevLett.104.116401}), where the long wavelength plasmon dispersion does not depend on the ultraviolet cutoff. The long wavelength plasmon dispersion varies linearly with the chemical potential $\mu$ or as $n^{1/3}$ with the electronic density. 

A few earlier works (including work from our group) report a slightly different version of the plasmon dispersion: $\omega_{\rm pl} = \omega_0$ in the $q \to 0$ limit \cite{PhysRevLett.102.206412,rashi_2015,anmol_2017}. This is a consequence of using the long wavelength approximation in the calculation of the (approximate) polarization function itself, which leads to vanishing overlap function in the inter-band contribution. This is technically incorrect since the overlap function contributes significantly in the inter-band part of the full polarization function, and leads  to the logarithmic terms in Eq.~\eqref{pl1}. Interestingly, in the very weak interaction limit ($\kappa \to \infty$), with $\alpha_{\rm ee}\ll1$, $\kappa^*(\omega) \to 1$ and Eq.~\eqref{pl1} reduces to $\omega_{\rm pl} = \omega_0$ to zeroth order in $q$, which is consistent with the expressions in earlier works\cite{PhysRevLett.102.206412,xiao_arxiv_2014,rashi_2015,anmol_2017}. 

Finally we note that for massless Dirac systems RPA is not exact even in the $q\to0$ limit, since the massless Dirac Hamiltonian of Eq.~\eqref{eqH} is not invariant to Galilean boosts, unlike the case of typical 2D/3D electron gas with parabolic dispersion \cite{2009arXiv0901.4528P,principipolini,PhysRevB.83.115135}. Thus, similar to the case of graphene, 
the RPA plasmon dispersion of Eq.~\eqref{pl1} can be expected to have some interaction induced renormalization correction, even in the long wavelength limit \cite{principipolini}. 
\subsection{Longitudinal optical conductivity}
Another observable connected to the longitudinal current current response function is the optical conductivity. 
The real part of long-wavelength longitudinal conductivity $\sigma(\omega)$, in the linear response regime,  is given by
\be\label{eq:22}
\text{Re}~\sigma(\omega)=-\frac{v_{\rm F}^2e^2}{\omega}\lim_{q\to0}\text{Im}\Pis(q\hatz,\omega)~.
\ee
Using Eq. (\ref{imagden2spin}) and Eq. (\ref{im13d}) in Eq. (\ref{eq:22}), the longitudinal conductivity for a single node of 3D Dirac semimetal can be obtained to be  
\be \label{eq:23}
\text{Re}~\sigma(\omega)=\frac{e^2\omega}{24\pi\hbar v_{\rm F}}~\Theta(\hbar\omega-2\mu)~,
\ee 
which is consistent with the results of Refs. [\onlinecite{PhysRevB.89.245121,PhysRevB.93.085426}]. Beyond the threshold of the Pauli blocked region, {\it i.e.}, for $\omega > 2 \mu$, the interband optical conductivity (involving only vertical transitions) is linearly proportional to the frequency in an ideal (very clean) 3D massless Dirac system. Such a linear dependence of the optical conductivity on the frequency has already been experimentally reported for ZrTe$_5$\cite{PhysRevB.92.075107}, in Cd$_3$As$_2$ with $001$ orientation\cite{neubauer_2016}, and in Eu$_2$Ir$_2$O$_7$ \cite{PhysRevB.92.241108}.
\section{Transverse spin-spin response function}
\label{Sec4}
To calculate the transverse spin-spin response function we use the expression as given in Eq. (\ref{response_func}) with ${\bf q} = q {\hat x}$ (in the $x-y$ plane)  which is perpendicular to the direction of the applied vector potential. Since the calculation from Eq.~\eqref{response_func}, proceeds in a manner similar to that of the density-density response function for which the results are well known, it is useful to express the transverse response functions also in terms of the density response function. Similar to the case of longitudinal response, we find that transverse response function (with ${\bf q}$ along the $\hat x$ direction), can also be expressed as a sum of 
intrinsic ($\mu =0$) as well as extrinsic ($\mu \neq 0$) contributions: 
\be
\Pis^{(0)}(q\hatx,\omega)=\Pis^{(0u)}(q\hatx,\omega)+\Pis^{(0d)}(q\hatx,\omega)~.
\ee
Let us consider the intrinsic (undoped) case first. 
\subsection{Undoped Case} 
To evaluate the transverse spin-spin response function, we follow an approach similar to that used for calculating the density-density response function in Ref.~[\onlinecite{xiao_arxiv_2014}]. The details of the calculations are presented in appendix~\ref{AppendixB}.
For the undoped case,  the imaginary and real component of transverse response function can be expressed in terms of the intrinsic density-density response function as 
\bea \label{transpi0}
\text{Im}\Pis^{(0u)}(q\hatx,\omega)&=&\frac{\omega^2-v_{\rm F}^2q^2}{v_{\rm F}^2q^2}\text{Im}\Pi_{\rho\rho}^{(0u)}(q,\omega)~, \\
\text{Re}\Pis^{(0u)}(q\hatx,\omega)&=&\frac{\omega^2-v_{\rm F}^2q^2}{v_{\rm F}^2q^2}\text{Re}\Pi_{\rho\rho}^{(0u)}(q,\omega)-\frac{ q_{\rm max}^2}{6\pi^2\hbar v_{\rm F}}~. \nn
\eea
It turns out that the relation between the density-density response and spin-spin response function in the undoped case, specified by Eq. (\ref{imagden2spin}-\ref{reden2spin}) for the longitudinal response and Eq.~\eqref{transpi0} for transverse response,  are identical to that  for massless Dirac systems in two dimension\cite{principipolini}. 

{Note that on account of gauge invariance, a real system cannot respond to static longitudinal vector potential. This implies that the longitudinal $\omega=0$ current-current response function should vanish  for every $q$ while the $\omega=0$ transverse response function should vanish for $q\to0$\cite{principipolini}. However Eq. (\ref{reden2spin}) and (\ref{transpi0}) do not satisfy this criteria in the static limit, on account of the presence of the ultraviolet cutoff dependent terms: $q_{\rm max}^2/(6\pi^2\hbar v_{\rm F})$. This is a direct consequence of the fact that gauge invariance of Eq.~\eqref{eqH} is explicitly broken by the ultraviolet energy cutoff. 
In order to restore the gauge invariance of the system, the static response function of the system should be corrected by subtracting the cut-off term $q_{\rm max}^2/(6\pi^2\hbar v_{\rm F})$ from Eqs.~(\ref{reden2spin}) and (\ref{transpi0}). This issue also arises in massive 2D Dirac systems\cite{Schliemann}, and can be cured by taking a lattice Hamiltonian (tight-binding) instead of the continuum Hamiltonian\cite{Stauber}. 

Similar divergences also arise while calculating the polarization bubble diagram in QED in 3+1 dimensions, with a cutoff regularization scheme. These are generally taken care of by adding a counter term in the Lagrangian to cancel such divergence. This leads to scale dependent and renormalized couplings constant: the Dirac quasiparticle charge in our case, and restores the gauge invariance of the low energy theory -- see the discussion following Eq.~\eqref{reden2spin} and Appendix \ref{QED}.

\subsection{Doped Case}
We now proceed to calculate the contributions to the transverse spin-spin response  function $\Pis^{(0d)}(q\hatx,\omega)$ at finite doping with $\mu > 0$. See appendix~\ref{AppendixB} for details.

The imaginary component of doped transverse spin-spin response function for a given $(q,\omega)$ is given by  
\be \label{im_tr3d}
\text{Im}~\Pis^{(0d)}=\frac{\omega^2-v_{\rm F}^2q^2}{32\pi\hbar v_{\rm F}^3q^3}
\begin{cases}
\beta(q,\omega)-\beta(q,-\omega), & \text{1A} \\
\beta(q,\omega), & \text{2A} \\
\frac{4q^3}{3}, & \text{1B} \\
\beta(-q,-\omega), & \text{2B} \\
0, & \text{3A}~ \\
0, & \text{3B}~. 
\end{cases}~ \\
\ee 
In Eq.~\eqref{im_tr3d}, the different regions are specified in Eq.~\eqref{eq:regions}/Fig.~(\ref{fig0}),  and we have defined the function: 
\be 
\beta(q,\omega)=2q\zeta(q,\omega)+q^2(2v_{\rm F}k_{\rm F}-v_{\rm F}q+\omega)~,
\ee  
where $\zeta(q,\omega)$ is defined in Eq. (\ref{eq:zeta}).

The real part of the doped component of the transverse spin-spin response function is a bit cumbersome. It can be expressed as
\bea \label{dopedtrans}
\text{Re}~\Pis^{(0d)}(q\hatx,\omega)&=&-\frac{\omega^2-v_{\rm F}^2q^2}{2v_{\rm F}^2q^2}~\text{Re}\Pi_{\rho\rho}^{(0d)}(q,\omega)\nn\\
&+&\frac{\omega^2-v_{\rm F}^2q^2}{16\pi^2\hbar v_{\rm F}^3q}\sum_{p=\pm1}[I_p^{(1)}(\omega)+I_p^{(1)}(-\omega)]\nn\\
&+&\frac{1}{16\pi^2\hbar v_{\rm F}q^3}\sum_{p=\pm1}I_p^{(2)}(q)~.
\eea 
 Here $I_p^{(1)}(\omega)$ and $I_p^{(2)}(q)$ are integrals defined as 
\bea
I_p^{(1)}&=&\int_{0}^{k_{\rm F}}dk\int_{l_1}^{l_2}dk' \frac{v_{\rm F}}{pv_{\rm F}k+\omega+v_{\rm F}k'}~,\\
I_p^{(2)}&=&\int_{0}^{k_{\rm F}}dk\int_{l_1}^{l_2}dk'(k'-pk)[(k'+pk)^2+q^2].
\eea
Here $l_1=|k-q|$, $l_2=|k+q|$. 
The exact analytic expression for these integrals are specified in Appendix-\ref{AppendixC}.

\subsection{Diamagnetic Susceptibility}
In this subsection, we use the obtained transverse current-current response function to calculate the diamagnetic susceptibility in 3D Dirac semimetals due to a static magnetic 
field. The noninteracting diamagnetic susceptibility $\Pi_{\rm orb}$ is given by\cite{giuliani},
\be \label{orbital}
\Pi_{\rm orb}=-\frac{v_{\rm F}^2e^2}{c^2} \lim_{q\to 0}\frac{\Pis^{(0)}(q\hatx,0)}{q^2}~,
\ee
where $\Pis^{(0)}(q\hatx,0)$ is the static transverse spin-spin response function. The static transverse susceptibility is purely real and it is given by 
\bea \label{static}
\Pis^{(0)}(q\hatx,0)&=&\frac{1}{48\pi^2q\hbar v_{\rm F}}\Bigg[ (4k_{\rm F}^3 + 3k_{\rm F} q^2)\log\left(\frac{|q-2k_{\rm F}|}{q+2k_{\rm F}}\right) \nn\\
&-&2q^3\log\frac{|4k_{\rm F}^2-q^2|}{4 q_{\rm max}^2} + \textcolor{blue}{4 k_{\rm F}^2 q } \Bigg]~. \nn \\
\eea                            
In the limit $q\to 0$, Eq. (\ref{static}) reduces to
\be
\lim_{q\to 0}\Pis^{(0)}(q\hatx ,0)=\frac{q^2}{12\pi^2\hbar v_{\rm F}}\log \left(\frac{\varepsilon_{\rm max}}{\mu}\right)~.
\ee
Substituting this in Eq. (\ref{orbital}) we obtain 
\be \label{orbital_sus}
\Pi_{\rm orb}=-\frac{v_{\rm F}^2e^2}{12\pi^2c^2\hbar v_{\rm F}} \log \left(\frac{\varepsilon_{\rm max}}{\mu}\right).
\ee
Note that since we have ignored the explicit structure of the Landau levels etc, the diamagnetic susceptibility obtained above 
should ideally hold in the weak field limit only. However it turns out that 
the diamagnetic susceptibility in Eq.~\eqref{orbital_sus} is consistent with a more rigorous calculation involving Landau levels, as it is consistent with the result of Eq.~(59) of Ref.~[\onlinecite{PhysRevB.81.195431}] in the $\Delta \to 0$ limit, and Eq.~(38) of Ref.~[\onlinecite{koshino-16}]. 

Similar anomalous divergence in diamagnetic susceptibility has been studied in detail in the context of Bi, which has an  anisotropic Dirac node \cite{Bismuth_Review}.  
A simple physical way of understanding the anomalous and diverging contribution is based on the energy argument. In a regular 
metal with parabolic dispersion, the Landau levels are equispaced in energy, thus the total energy of the system (for levels away from the chemical potential) remains unchanged on the application of the magnetic field. Only the states in vicinity of the chemical potential are affected and give a finite contribution to the diamagnetic susceptibility. However in systems with a Dirac node, the Landau-levels are not equispaced, and as a consequence the total energy of the system gets finite contribution from all filled states (primarily the infinite levels in the valance band), and thus the orbital/diamagnetic susceptibility diverges. This argument can also be quantified based on a crude estimate by second order perturbation theory \cite{Moll_torque}. The external magnetic field $B$ couples to the orbital degrees of freedom via the vector potential $A \approx B \times r \approx B/k$. Thus the correction to the total energy of the system upto second order in $A$ is (for a given {\bf k} mode) is 
$ \approx (B/k)^2/(v_{\rm F} k)$, where $2v_{\rm F}k$ is the transition energy from valance to conduction band. Summing over the allowed ${\bf k}$ modes now yields, $\delta E(B) \propto  B^2 \log(\varepsilon_{\max}/\mu)$, consistent with Eq.~\eqref{orbital_sus}. This crude estimate is also consistent with a more thorough calculation for the change in the total energy of a Dirac node in presence of a magnetic field \cite{Moll_torque}.


\section{Current-Current response in Weyl semimetals with chiral anomaly}
\label{Sec5}
Having calculated the current-current (and spin-spin) response functions for a single Dirac node, we now proceed to calculate the corresponding response for a Weyl semimetal. {For simplicity, we consider a Weyl semimetal with one pair of Weyl nodes. 
For an experimental system with $g$ pairs of Weyl nodes, all our results should be multiplied by $g$. 
The low energy Hamiltonian, for each of the  Weyl node of chirality $\chi = \pm 1$, in vicinity of the Weyl node, is 
\be \label{Hweyl}
{\mathcal{H}}_{\chi}=\hbar v_{\rm F}~\chi~{\bf k} \cdot {\bm \sigma} - \mathbb{I} \mu_{\chi}
\ee
where $\bm \sigma$ is the vector of the three Pauli matrices and $\mathbb{I}$ is the $2 \times 2$ identity matrix and $\mu_{\chi}$ denotes
the chemical potential of the Weyl node of chirality $\chi$. 

Application of parallel electric (${\bf E}$) and magnetic (${\bf B}$) fields  in a Weyl semimetal leads to a chiral anomaly: charge transfer from the $\chi = -1$ node to the $\chi = +1$ node for ${\bf E}\cdot{\bf B} >0$ and vice versa for ${\bf E}\cdot{\bf B} < 0$. This charge transfer is eventually stabilized by some inter-node scattering mechanism with timescale $\tau$. 
The amount of electron transferred from one Weyl node to other is given by $\Delta n=   \frac{e^2}{2\pi^2\hbar^2}{\bf E\cdot B}~\tau$. This leads to a shift in the respective chemical potentials in the two Weyl nodes. If initially both the nodes were doped with a chemical potential $\mu = \hbar v_{\rm F} k_{\rm F}$, with $k_{\rm F}^3 = 6 \pi^2 n$, then the modified densities in the two nodes with $\chi = \pm 1$ are $n_{\pm} = n \pm \Delta n/2$. Accordingly their modified Fermi wave-vectors are given by $(k_{\rm F\pm})^3 = 6 \pi^2 n_{\pm}$, and the corresponding chemical potential is given by \cite{xiao_arxiv_2014}, 
\be\label{muchiral}
\mu_{\pm}=\left(\mu^3 \pm \frac{3e^2\hbar v_{\rm F}^3}{2}{\bf E\cdot B}~\tau \right)^{1/3}.
\ee
For the rest of the article, we will be working in the weak magnetic field limit, whereby the discrete Landau levels structure of the Weyl semimetal can be ignored. Additionally we assume that ${\bf E} \parallel {\bf B}$ (or ${\bf E}\cdot{\bf B} >0$) and consequently we have $\mu_+ > \mu_-$. 

Equation \eqref{muchiral} implies that for a physical manifestation of the chiral anomaly to be seen in experiments, the splitting of the chemical potential should be of the order of the chemical potential. This implies that $|\mu_+ - \mu_-| \approx \mu$ or alternately to lowest order in ${\bf E}\cdot{\bf B}$ term, $ e^2\hbar v_{\rm F}^3 {\bf E\cdot B}~\tau \approx \mu^3$. Typically the inter-valley scattering time is of the order of $\tau \approx 10^{-9}$s with the corresponding length scale of the order of a few microns - see Ref.~[\onlinecite{PhysRevX.4.031035}], for a discussion. Assuming $v_{\rm F}$ to be of the order of $10^{6}$m/s, this implies that ${\bf E \cdot B} \approx \mu^3 \times 10^6$ Tesla V/m, where $\mu$ is expressed in eV.  Thus if  $\mu \approx 0.1$ eV, then ${\bf E \cdot B}$ should be of the order of $10^3$ Tesla V/m for the chiral anomaly to be distinguishable in experiments.

%
\begin{figure}[t] 
\begin{center} 
\includegraphics[width=0.9\linewidth]{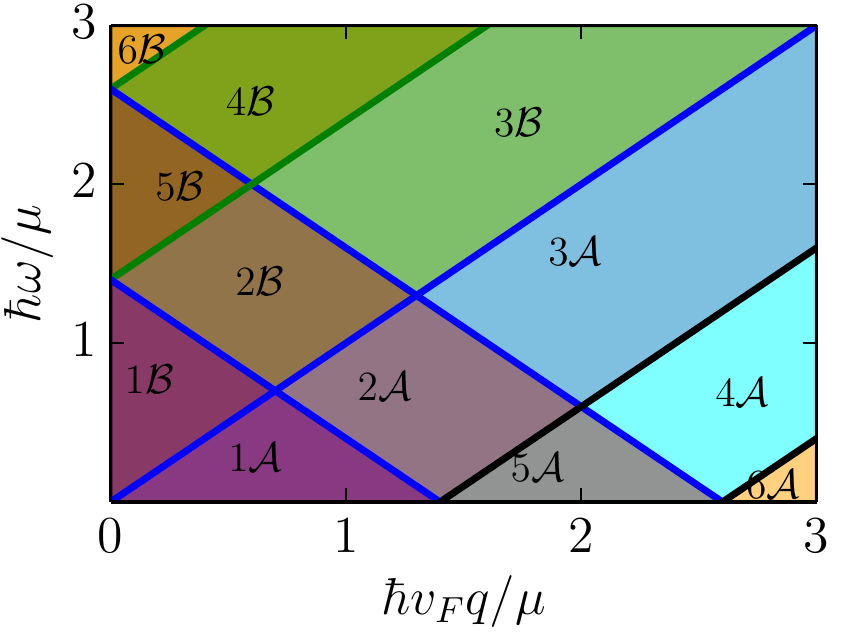} 
\end{center} 
\caption{Different regions in the $\omega -q$  plane used to define the imaginary part of the longitudinal as well as the  transverse spin-spin response function in a Weyl semimetal with chiral anomaly. Here the regions $2\mathcal{B}$, $3\mathcal{B}$
and $4\mathcal{B}$ have interband single particle excitations while the regions  $1\mathcal{A}$, $2\mathcal{A}$, $3\mathcal{A}$,
$4\mathcal{A}$ and $5\mathcal{A}$ have intraband
single particle excitations for either one or both the Weyl nodes with Fermi energy located at $\mu_{\pm}$. 
More specifically the $3\mathcal{B}$ region has interband particle-hole excitation of both nodes, and 
the $3\mathcal{A}$ has intraband particle-hole excitations of both the Weyl nodes. 
The regions $1\mathcal{B}$, $5\mathcal{B}$, $6\mathcal{B}$ and
$6\mathcal{A}$ are without any single particle excitation. 
} 
\label{fig1}  
\end{figure}
%

The longitudinal and the transverse spin-spin response functions for a Weyl semimetal can now be obtained simply by summing the contributions from the two Weyl nodes with the modified chemical potential \cite{xiao_arxiv_2014,PhysRevB.91.081106}. 
\subsection{Longitudinal response}
The imaginary part for longitudinal spin-spin response function at finite doping modifies to 
 \bea \label{im13dchiral}
 &&\text{Im}~\Pis^{(0d)}(q\hatz,\omega)=
 -\frac{\omega^2}{8\pi\hbar v_{\rm F}^3q^2} \\
 &&\times 
 \begin{cases}
 \sum_{\chi = \pm1} \left[\zeta_{(\chi)}(q,\omega)-\zeta_{(\chi)}(q,-\omega)\right] & 1\mathcal{A}\\
 \zeta_{+}(q,\omega)-\zeta_{+}(q,-\omega) 
 +\zeta_{-}(q,\omega), & 2\mathcal{A} \\
 \zeta_{+}(q,\omega)+\zeta_{-}(q,\omega), & 3\mathcal{A} \\
 \zeta_{+}(q,\omega), & 4\mathcal{A} \\
 \zeta_{+}(q,\omega)-\zeta_{+}(q,-\omega), & 5\mathcal{A} \\
 -\frac{2q^2}{3}, & 1\mathcal{B}\\
 -\zeta_{-}(-q,-\omega)-\frac{q^2}{3}, & 2\mathcal{B}\\
 -\zeta_{+}(-q,-\omega)-\zeta_{-}(-q,-\omega), & 3\mathcal{B}\\
 -\zeta_{+}(-q,-\omega), & 4\mathcal{B}\\
 -\frac{q^2}{3}, & 5\mathcal{B} \\
 0, & 6\mathcal{A},~6\mathcal{B}.\\
 \end{cases}~
 \eea 
The different regions defined above can be evaluated by replacing $\mu\to\mu_\chi$ in Eq. (\ref{eq:regions}) and are also displayed in Fig. (\ref{fig1}). Additionally we have defined the function, 
\be \label{zetapm}
\zeta_{\pm}=\frac{1}{12\hbar^3 v_{\rm F}^3q}\left[(2\mu'_\pm)^3-3\hbar^2 v_{\rm F}^2q^2(2\mu'_\pm)+2\hbar^3 v_{\rm F}^3q^3\right]~, 
\ee 
with $2\mu'_{\pm} = 2\mu_\pm+\hbar\omega$.

Following a similar procedure, the real part of longitudinal spin-spin response function of a Weyl semimetal with 2 nodes can be expressed as 
\be
 \text{Re}~\Pis^{(0d)}(q\hatz,\omega)=-\frac{\omega^2}{8\pi^2\hbar v_{\rm F}^3}\sum_{\chi=\pm1}[\mathcal{C}_\chi(q,\omega)+\mathcal{D}_\chi(q,\omega)]~.
\ee 
Here we have defined the functions,
\bea
 \mathcal{C}_{\chi}(q,\omega) & =& \mathcal{C}(q,\omega)|_{\mu \to \mu_\chi}~, \nn \\ 
 \mathcal{D}_{\chi}(q,\omega) &=& \mathcal{D}(q,\omega)|_{\mu \to \mu_\chi}~, \nn 
\eea
which in turn use the definitions from Eq.~\eqref{eq:CD}.
 
Substituting the above expression in Eq. (\ref{pl1}), the plasmon dispersion in long-wavelength limit upto order of $q^2$ modifies to
\bea \label{plchiral}
&&\omega_{\rm pl}=\omega_0 \left[1-\frac{\hbar^2v_{\rm F}^2q^2}{8[(\mu_+)^2+(\mu_-)^2]}\sum_{\chi=\pm1}\left\{1+\gamma(\hbar\omega/2\mu_\chi)\right\}\right]~.\nn\\
\eea
In Eq. (\ref{plchiral}) we have used,
\be \label{eq:omega0}
\hbar\omega_0=\sqrt{\frac{2\alpha_{\rm ee}[(\mu_+)^2+(\mu_-)^2]}{3\pi\kappa^{*}(\omega_0)} }~.
\ee
and the effective frequency dependent background dielectric constant is given by
\be
\kappa^{*}(\omega)= 1+\frac{\alpha_{\rm ee}}{6\pi}\sum_{\chi=\pm1}\log\left|\frac{4\hbar^2v_{\rm F}^2q_{\rm max}^2}{\hbar^2\omega^2-4(\mu_\chi)^2}\right|~.
\ee
In the absence of chiral anomaly, $|\mu^+|=|\mu^-|$ and Eq.~\eqref{plchiral} reduces to Eq. (\ref{pl1}). The results for the plasmon dispersion in Eq.~\eqref{plchiral} are consistent with those derived in Ref.~[\onlinecite{xiao_arxiv_2014}].
Substituting the value of $\mu_{\pm}$ from Eq. (\ref{muchiral}) in \eqref{eq:omega0}, to leading order in ${\bf E}\cdot{\bf B}$, $\omega_0$ 
can be expressed in terms of the external parallel electric and magnetic field as 
\be
\hbar\omega_0=\mu\sqrt{\frac{4 \alpha_{\rm ee}}{3\pi\kappa^{*}(\omega_0)}}\left(1-\frac{\hbar^2 e^4v_{\rm F}^6\tau^2({\bf E\cdot B})^2}{8\mu^6}\right)~. 
\ee 
Thus the $q\to 0$ plasmon dispersion is expected to have a dependence of $({\bf E}\cdot{\bf B})^2$ to leading order in ${\bf E}$ and ${\bf B}$ fields \cite{xiao_arxiv_2014}.

Another experimental observable which carries signature of the chiral anomaly is the optical conductivity. In presence of the chiral anomaly, the longitudinal conductivity defined in Eq.~\eqref{eq:22} gets modified and it is given by \cite{PhysRevB.89.245121},
\be \label{con_chiral}
{\rm Re}~\sigma(\omega)=\frac{e^2\omega}{24\pi\hbar v_{\rm F}}~\left[\Theta(\hbar\omega-2\mu_+)+\Theta(\hbar\omega-2\mu_{-})\right]~.
\ee 
The additional step function appearing in Eq.~(\ref{con_chiral}) as compared to Eq.~(\ref{eq:23}) is a consequence of different Pauli blocking of the optically excited carriers in the two Weyl nodes due to different chemical potential in presence of a chiral anomaly. 
In absence of chiral anomaly, there is linear optical conductivity beyond the chemical potential as a function of $\omega$, while in presence of a chiral anomaly an extra step function with linear dependence on $\omega$ appears in the optical conductivity, with the width of the step function being proportional to ${\bf E}\cdot{\bf B}$\cite{PhysRevB.89.245121}.

\subsection{Transverse response}
In this subsection we explore the impact of chiral anomaly on the transverse current-current response function, and the orbital susceptibility. The imaginary component of the transverse spin-spin response function is given by, 
\bea \label{im13dchiral2}
&&\text{Im}~\Pis^{(0d)}(q\hatx,\omega)=
\frac{\omega^2-v_{\rm F}^2q^2}{32\pi\hbar v_{\rm F}^3q^3}\\
&&\times 
\begin{cases}
\sum_{\chi = \pm} \beta_{\chi}(q,\omega)-\beta_{\chi}(q,-\omega),  & 1\mathcal{A} \\
\beta_{+}(q,\omega)-\beta_{+}(q,-\omega) +\beta_{-}(q,\omega), & 2\mathcal{A} \\
\beta_{+}(q,\omega)+\beta_{-}(q,\omega), & 3\mathcal{A} \\
\beta_{+}(q,\omega), & 4\mathcal{A} \\
\beta_{+}(q,\omega)-\beta_{+}(q,-\omega), & 5\mathcal{A} \\
\frac{8q^3}{3}, & 1\mathcal{B}\\
\beta_{-}(-q,-\omega)+\frac{4q^3}{3}, & 2\mathcal{B}\\
\beta_{+}(-q,-\omega)+\beta_{-}(-q,-\omega), & 3\mathcal{B}\\
\beta_{+}(-q,-\omega), & 4\mathcal{B}\\
\frac{4q^3}{3}, & 5\mathcal{B} \\
0, & 6\mathcal{A},~6\mathcal{B}~,\\ 
\end{cases}~\nn
\eea 
where the different regions in $\omega-q$ plane are shown in Fig. (\ref{fig1}) and we have defined the function, 
\be
\beta_{\pm}(q,\omega)=2q~\zeta_\pm(q,\omega)+q^2(2\mu_\pm+\hbar\omega-\hbar v_{\rm F}q)~.
\ee

The real component of the doped transverse spin-spin response function is explicitly given by 
\bea
&&{\rm Re}~\Pis^{(0d)}(q\hatx,\omega)=\frac{\omega^2-v_{\rm F}^2q^2}{16\pi^2\hbar v_{\rm F}^3}\sum_{\chi=\pm1}[\mathcal{C}_\chi(q,\omega)+\mathcal{D}_\chi(q,\omega)]\nn\\
&&+\frac{\omega^2-v_{\rm F}^2q^2}{16\pi^2\hbar v_{\rm F}^3q}\sum_{p=\pm1}\sum_{\chi=\pm1}\Big[I_p^{(1)}(k_{{\rm F}\chi},\omega)+I_p^{(1)}(k_{{\rm F}\chi},-\omega)\Big]\nn \\
&&+\frac{1}{16\pi^2\hbar v_{\rm F}q^3}\sum_{p=\pm1}\sum_{\chi=\pm1}~I_p^{(2)}(k_{{\rm F}\chi},q)~.
\eea  
Here we have defined the chiral Fermi wavevector $k_{{\rm F}\chi} = \mu_\chi/(\hbar v_F)$ and the explicit form of the integrals $I_p^{(1)}$ and $I_p^{(2)}$ are specified in Appendix-\ref{AppendixC}.

The static part of transverse spin-spin response function of a Weyl semimetal with chiral anomaly modifies as
\bea\label{stat_chiral}
&&\Pis^{(0)}(q\hatx,0)=\frac{1}{48\pi^2\hbar v_{\rm F}q}\sum_{\chi=\pm1}\Bigg[2q^3\log\left(\frac{4 q_{\rm max}^2}{|4k_{{\rm F}\chi}^2-q^2|}\right)\nn\\
&+&[4k_{{\rm F}\chi}^3+3k_{{\rm F}\chi}q^2]\log\left(\frac{|q-2k_{{\rm F}\chi} |}{q+2k_{{\rm F}\chi}}\right)
{+4k_{\rm F \chi}^2q}\Bigg]~.
\eea
Using Eq. (\ref{stat_chiral}) to evaluate the diamagnetic susceptibility, we obtain
\be\label{orbital_chiral} 
\Pi_{\rm orb}=-\frac{v_{\rm F}^2e^2}{12\pi^2c^2\hbar v_{\rm F}}\log\left(\frac{\varepsilon_{\rm max}^2}{|\mu_{-}||\mu_{+}|}\right) ~.
\ee
To leading order in $({\bf E}\cdot{\bf B})$, 
the diamagnetic susceptibility is  
\be
\Pi_{\rm orb}=-\frac{v_{\rm F}^2e^2}{12\pi^2c^2\hbar v_{\rm F}}\left[2\log\frac{\varepsilon_{\rm max}}{\mu}+\frac{3\hbar^2e^4v_{\rm F}^6\tau^2({\bf E\cdot B})^2}{4\mu^6}\right]~.
\ee

\section{Implications for anisotropic Dirac and Weyl systems} 
\label{new}
While we have focussed primarily on isotropic systems, several actual experimental realization of  Dirac/Weyl semimetals such as Na$_3$Bi \cite{DSM_A3Bi,Liu864,PhysRevB.94.085121}, Cd$_3$As$_2$ \cite{Cd3As2_1,Cd3As2_Natmat,DSM_Cd3As2,PhysRevLett.113.027603,Cd3As2_2}, PtTe$_2$\cite{PtTe_2} etc., host anisotropic Dirac/Weyl fermions. Thus in this section we qualitatively discuss the implications of our isotropic calculations for systems with anisotropic dispersion.

Let the anisotropic velocities for of the anisotropic Dirac cone be $\{v_x,v_y,v_z\}$. The anisotropy in the response functions is likely to be captured by the following replacement: $q \to q'$ where 
\be 
q' = q \left[{\sin^2}\theta_q \left( \cos^2\phi_q + \frac{v_y^2}{v_x^2} \sin^2\phi_q \right) + \frac{v_z^2}{v_x^2} \cos^2\theta_q~ \right]^{1/2}, 
\ee
and $v_F \to v_x$, in all the response functions. A similar replacement was shown to arise in the calculation of the density density response function for an anisotropic and tilted Dirac cone in two dimensions, in the context of Borophene \cite{PhysRevB.96.035410}.
Accordingly a similar direction dependent anisotropy factor will also appear in the plasmon dispersion. However the qualitative features of the gapped plasmon dispersion with $\omega_{\rm pl} \propto \mu$  or $\omega_{\rm pl} \propto n^{1/3}$ should not change. 

As far as the optical conductivity (longitudinal response) of a given Dirac node is concerned, it should be still given by a form similar to Eq.~\eqref{eq:23}, with the Fermi velocity being substituted\cite{PhysRevB.92.241108} by $v_F \to {(v_x v_y v_z)^{1/3}}$.  
For the diamagnetic magnetic susceptibility (transverse response) also we expect the same functional form as Eq.~\eqref{orbital_sus}, with the substitution of $v_F \to v_z$. The qualitative behaviour involving logarithmic divergence in the orbital susceptibility, i.e.,  $\Pi_{\rm orb} \propto \log \left({\varepsilon_{\rm max}}/{\mu}\right)$, should persist even in the anisotropic case.

\section{Summary}
\label{Sec6}
To summarize,  we have presented the analytical results for the longitudinal and transverse 
current-current response function in the $(\omega,q)$ plane for a single Dirac/Weyl node. 
As expected, the current-current response function is related to the density-density response function due to the 
charge continuity equation. Additionally since the current operator is proportional to the spin operator, the current-current 
response function is also related to the spin-spin response function. 

We find that for the undoped 3D Dirac node, 
the relation between spin-spin response function and density-density response function are identical to that for a for 
2D Dirac node as in graphene. However for the case of finite doping the relationship 
between the response functions differ between 2D and 3D. For a 3D Dirac node, the long wavelength plasmon dispersion is directly proportional to the chemical potential, the optical conductivity beyond the Pauli blocked regime is linearly proportional to the frequency and the diamagnetic susceptibility diverges logarithmically for vanishing chemical potential.  

The current current response function of a single Dirac node is then used to obtain the response function of a Weyl Semimetal with chiral anomaly. For a Weyl semimetal in presence of parallel ${\bf E}$ and ${\bf B}$ fields, we find that the long wavelength plasmon dispersion (or gap) is proportional to $({\bf E}\cdot{\bf B})^2$ to leading order in ${\bf E}\cdot{\bf B}$, the optical conductivity displays a two step behaviour due to partial Pauli blocking in one of the Weyl nodes, and the diamagnetic susceptibility is found to vary as  $({\bf E}\cdot{\bf B})^2$ to leading order in ${\bf E}\cdot{\bf B}$. 

\section{Acknowledgements}
AA thanks the INSPIRE faculty fellowship award by the Dept. of Science and Technology, Government of India, for the financial support. We also thank Joydeep Chakrabortty for stimulating discussions. 

\appendix
\section{The commutator term}
\label{AppendixA}
Here we evaluate the commutator term appearing in Eq.~(\ref{den2spin}). A similar calculation has already been done for the 2D case of graphene in Ref.~[\onlinecite{PhysRevB.78.075410}] and we follow a similar approach here. 
For massless electrons described by unbounded linear relations the operators are also defined in the unbounded energy/momentum space. For unbounded operators in general $\sum_{k}\langle{\cal{O}}(k)\rangle \neq \sum_{k+q}\langle{\cal{O}}(k+q)\rangle$. To overcome this, one defines `bounded' operators by subtracting out the ground state contribution \cite{PhysRevB.78.075410}. This is done by defining normal ordered operators:
\be
:G(k):=G(k)- \langle 0|G(k)|0\rangle~.
\ee

Expressing the commutator defined in Eq. (\ref{den2spin}) in terms of normal ordered operators yields, 
\bea  \label{app1}
\big[{\bf q}.{\bf J}_q,\rho_{-q}\big] = -v_{\rm F}\sum_{\bf k} &\Big[&\langle 0|\psi_{\bf{k-q}}^\dagger({\bf q.\sigma})\psi_{{\bf k-q}}|0\rangle \nn \\
&-&\langle 0|\psi_{\bf{k}}^\dagger({\bf q.\sigma})\psi_{{\bf k}}|0\rangle\Big]~,
\eea
where the ground state comprises of the completely filled valence band. However for doped Weyl semimetals with $\mu > 0$, the contributions from electron above the Dirac points also have to be accounted for. The difference of the two infinite sums in Eq.~ (\ref{app1}), can be computed by regularizing it with the high energy ultraviolet cutoff. To proceed further, we switch to a diagonal basis by diagonalizing the matrices defined in Eq. (\ref{app1}) by using the unitary transformation that diagonalizes the Hamiltonian:
\be
U_{\bf{k}}=\left(\begin{array}{cc}
\cos\frac{\theta_k}{2}e^{-i\phi_k/2} & \sin\frac{\theta_k}{2}e^{i\phi_k/2}\\ 
\sin\frac{\theta_k}{2}e^{-i\phi_k/2} & -\cos\frac{\theta_k}{2}e^{i\phi_k/2} \end{array} \right)~.
\ee

For the longitudinal case, taking ${\bf q}$ along the $z$ direction Eq. (\ref{app1}) simplifies to
\be
\big[\hat{j}_{z,q},\rho_{-q\hat{{\bf z}}}\big] = v_{\rm F}\sum_{\bf k}[\cos\theta_{\bf{k+q}}-\cos\theta_{\bf{k}}]~.
\ee
For evaluating the sum, the summation over $\bf k$ is converted to the integral in the $k$ space using the ultraviolet cutoff $q_{\rm max}$ for the maximum allowed $k$:
\be \label{eq:sum}
\sum_{k}[\cos\theta_{\bf{k+q}}-\cos\theta_{\bf{k}}]=\frac{1}{(2\pi)^3}\int_{k=0}^{k_{\text{max}}} k^2 \cos\theta~d\Omega~,
\ee
where $k_{\text{max}}=\sqrt{q^2+q_{\rm max}^2+2q_{\rm max} q \cos\theta}$ and $d\Omega=\sin\theta~d\theta~d\phi $ is the integration over the solid angle. Note that in the l.h.s. of Eq.~\eqref{eq:sum}, only the first term contributes to the integral and we have chosen ${\bf q}$ along the $\hat{z}$ direction. Taking the leading order contribution in $q_{\text{max}}$ on the r.h.s. of Eq.~\eqref{eq:sum} we get 
\be
\frac{1}{v_{\rm F}q}\big[\hat{j}_{z,q},\rho_{-q\hat{{\bf z}}}\big]=\frac{ q_{\rm max}^2}{6\pi^2}~.
\ee
\section{The longitudinal and transverse overlap function} 
Here we briefly discuss the calculation of the overlap function appearing in our calculations. 
The eigenfunction of the Hamiltonian in Eq.~\eqref{eqH} are given by,
\be \label{ol}
\chi_\lambda({\bf k})=\left(\begin{array}{c} e^{i\phi_{\bf k}} \cos{(\theta_{\lambda,{\bf k}}/{2})} \\ \lambda\sin{(\theta_{\lambda,{\bf k}}/{2})}~
\end{array}\right)~.
\ee
Here $\lambda=\pm1$ denotes the conduction and valence band respectively and $\theta_{\lambda, {\bf k}}=\theta_{\bf k}$ for $\lambda=1$ and $\pi-\theta_{\bf k}$ for $\lambda=-1$. Here $\theta_{\bf k}$ and $\phi_{\bf k}$ are simply the 
angles related to the point ${\bf k}$ in spherical coordinates, with $\cot \theta_{\bf k} = k_z/\sqrt{k_x^2+k_y^2}$ and $\tan \phi_{\bf k} = k_y/k_x$. 
Using Eq.~\eqref{ol}, the overlap function can be evaluated to be,
\bea
& & f^{\lambda\lambda'}({\bf k},{\bf k'})=|\langle\chi_\lambda({\bf k})|\sigma_z|\chi_{\lambda'}({\bf k'})\rangle|^2\\
& & =\frac{1}{2}[1+\lambda\lambda'(\cos\theta_{\bf k}\cos\theta_{\bf k'}-\sin\theta_{\bf k}\sin\theta_{{\bf k}'}
\cos\phi_{{\bf k}{\bf k}'})]\nn.
\eea
Here ${\bf k'}={\bf k}+{\bf q}$ and $\phi_{{\bf k}{\bf k}'}$ is the angle between ${\bf k'}$ and ${\bf k}$. 

Specifically for the longitudinal case we have ${\bf q}=q\hat{z}$ and the longitudinal overlap function is given by,
\be
f_L^{\lambda\lambda'}({\bf k},{\bf k'})=\frac{1}{2}\left[1+\lambda\lambda'\frac{k\cos2\theta_{\bf k}+ q\cos\theta_{\bf k}}{k'}\right]~.
\ee
For the transverse case where we have chosen ${\bf q}=q\hat{x}$, and accordingly we have 
\be
f_T^{\lambda\lambda'}({\bf k},{\bf k'})=\frac{1}{2}\left[1+\lambda\lambda'\frac{k\cos2\theta_{\bf k}- q\sin\theta_{\bf k}\cos\phi_{\bf k}}{k'}\right]~.
\ee  
\section{The transverse spin-spin response function}\label{AppendixB}
\begin{widetext}
The transverse spin-spin response function can expressed as a sum of the intrinsic (undoped) and extrinsic (doped) part:
\be \label{ap2eq1}
\Pis^{(0)}(q\hatx,\omega)=\Pis^{(0u)}(q\hatx,\omega)+\Pis^{(0d)}(q\hatx,\omega)~.
\ee 
The extrinsic contribution is further decomposed into inter-band and intra-band transitions as
\be \label{ap2eq2}
 \Pis^{(0d)}(q\hatx,\omega)=\chi_{k_{\rm F}}^{-}(q\hatx,\omega)+\chi_{k_{\rm F}}^{+}(q\hatx,\omega)~.
\ee 
The functions, $\chi_{k_{\rm F}}^{-}(q\hatx,\omega)$ and $\chi_{k_{\rm F}}^{+}(q\hatx,\omega)$ are defined as 
\bea
\chi_{k_{\rm F}}^{-}(q\hatx,\omega)&=&-\frac{1}{L^3}\sum_{{\bf k}<k_{\rm F}}f^{-}_T({\bf k},{\bf k+q})\left(\frac{1}{\hbar \omega -\varepsilon_{\bf k}-\varepsilon_{\bf k+q}+i\epsilon}-\frac{1}{\hbar \omega +\varepsilon_{\bf k}+\varepsilon_{\bf k+q}+i\epsilon}\right)~, \label{ap2eq4}\\
\chi_{k_{\rm F}}^{+}(q\hatx,\omega)&=&\frac{1}{L^3}\sum_{{\bf k}<k_{\rm F}}f^{+}_T({\bf k},{\bf k+q})\left(\frac{1}{\hbar \omega +\varepsilon_{\bf k}-\varepsilon_{\bf k+q}+i\epsilon}-\frac{1}{\hbar \omega -\varepsilon_{\bf k}+\varepsilon_{\bf k+q}+i\epsilon}\right)~,\label{ap2eq5}
\eea 
where $\varepsilon_{\bf k}=\hbar v_{\rm F} |{\bf k}|$ and $f^{\pm}_T({\bf k},{\bf k+q})$ is the band overlap function of normalized eigen spinors. In terms of these $\Pis^{(0u)}(q\hatx,\omega) = - \chi_{q_{\rm max}}^{-}(q\hatx,\omega)$. 
Defining  $\varepsilon_{\bf k+q}/(\hbar v_{\rm F})=k'$, and $\omega'=({\omega}+i\epsilon)/v_{\rm F}$,  
the overlap function is given by 
\be \label{ap2eq6}
f^{\pm}_T({\bf k},{\bf k+q})=\mp\frac{[(k'\mp k)^2-q^2][q^2\sin^2\phi+(k'\pm k)^2\cos^2\phi]}{4kk'q^2}~.
\ee  
Converting the sum into integrals, Eqs.~\eqref{ap2eq4}-~\eqref{ap2eq5}, reduce to the following:
\bea
&&\chi_{k_{\rm F}}^{-}(q\hatx,\omega)=-\frac{1}{32\pi^2\hbar v_{\rm F}q^3}\int_{0}^{k_{\rm F}}dk\int_{l_1}^{l_2} dk' [(k'+k)^2-q^2][q^2+(k'-k)^2]\left(\frac{1}{\omega'-k-k'}\right)+\omega'\to-\omega'~,\label{ap2eq8}\\
&&\chi_{k_{\rm F}}^{+}(q\hatx,\omega)=-\frac{1}{32\pi^2\hbar v_{\rm F}q^3}\int_{0}^{k_{\rm F}}dk\int_{l_1}^{l_2} dk' [(k'-k)^2-q^2][q^2+(k'+k)^2]\left(\frac{1}{\omega'+k-k'}\right)+\omega'\to-\omega'~,\label{ap2eq9}
\eea 
where $l_1=|k-q|$ and $l_2=|k+q|$. Focussing on only $\omega'>0$ case, we first evaluate the intrinsic contribution $\Pis^{(0u)}(q\hatx,\omega)$. Using the \textit{Sokhotski~Plemelj~theorem}: $1/{(x\pm i\epsilon)}=\mathbb{P}({1}/{x})\mp i\pi\delta(x)$, with $\mathbb{P}$ denoting the principal value of the integral, 
we obtain 
\bea 
&&\text{Im}~\Pis^{(0u)}(q\hatx,\omega)=-\frac{1}{32\pi\hbar v_{\rm F}q^3}\int_{0}^{q_{\rm max}}dk\int_{l_1}^{l_2} dk' [(k'+k)^2-q^2][q^2+(k'-k)^2]\delta(\tilde{\omega}-k-k') \label{ap2eq10}~,\\
&&\text{Re}~\Pis^{(0u)}(q\hatx,\omega)=\frac{1}{32\pi^2\hbar v_{\rm F}q^3}\mathbb{P}\int_{0}^{q_{\rm max}}dk\int_{l_1}^{l_2} dk' [(k'+k)^2-q^2][q^2+(k'-k)^2]\left(\frac{1}{\tilde{\omega}-k-k'}-\frac{1}{\tilde{\omega}+k+k'}\right).\label{ap2eq11}
\eea 
Evaluating the integrals in Eq. (\ref{ap2eq10})-(\ref{ap2eq11}) yields, 
\bea 
&&\text{Im}~\Pis^{(0u)}(q\hatx,\omega)=-\frac{\omega^2-v_{\rm F}^2q^2}{24\pi\hbar v_{\rm F}^3}\Theta(\omega-q)\\
&&\text{Re}~\Pis^{(0u)}(q\hatx,\omega)=-\frac{\omega^2-v_{\rm F}^2q^2}{24\pi\hbar v_{\rm F}^3}\log \left(\frac{4v_{\rm F}^2q_{\rm max}^2}{|v_{\rm F}^2q^2-\omega^2|}\right)-\frac{q_{\rm max}^2}{6\pi^2\hbar v_{\rm F}}
\eea  
Following the similar procedure one can calculate the imaginary and real component of the doped transverse spin-spin response function, by integrating Eq.~\eqref{ap2eq8} and \eqref{ap2eq9}.

\section{Integrals $I_p^{(1)}(\omega)$ and $I_p^{(2)}(q)$}\label{AppendixC}
The expression in Eq. (\ref{dopedtrans}) for real component of doped transverse spin-spin response function involves the integrals defined as 
\bea 
I_p^{(1)}(k_{\rm F},{\tilde{\omega}})&=&\int_{0}^{k_{\rm F}}dk\int_{l_1}^{l_2} \frac{dk'}{pk+k'+{\tilde{\omega}}}~,\label{c1}\\
I_p^{(2)}(k_{\rm F},q)&=&\int_{0}^{k_{\rm F}}dk\int_{l_1}^{l_2}dk'(k'-pk)[(k'+pk)^2+q^2]~,\label{c2}
\eea
where $l_1=|k-q|$, $l_2=|k+q|$ and we have defined ${\tilde{\omega}}=\omega/v_{\rm F}$. Doing the integral over $k'$ and $k$ we obtain
\bea \label{Ip1}
I_p^{(1)}(k_{\rm F},{\tilde{\omega}})&=&\Bigg[\eta(k_{\rm F},q,p-1,q+{\tilde{\omega}})-\eta(k_{\rm F},q,p+1,-q+{\tilde{\omega}})\Bigg]~\Theta(k_{\rm F}-q)\nn\\
&& + \Bigg[\eta(k_{\rm F},0,p+1,q+\tilde{\omega})-\eta(k_{\rm F},0,p-1,q+\tilde{\omega})\Bigg]~.
\eea
In Eq.~\eqref{Ip1}, we have defined the function 
\bea
\eta(a,b,c,d) \equiv \int_{b}^{a} \log |ck+d|~dk = \frac{1}{c}\left[(b-a)c+(ac+d)\log(|ac+d|)-(bc+d)\log(|bc+d|)\right]~.
\eea

For the other integral, performing the integration over $k'$ and $k$ in Eq. (\ref{c2}), gives 
\bea
I_p^{(2)}(k_{\rm F},q)&=&\left(2k_{\rm F}^2q^3-\frac{4}{15}k_{\rm F}^5p\right)+\frac{2p}{15}\Bigg[2k_{\rm F}^5-5k_{\rm F}^2q^3+3q^5\Bigg]\Theta(k_{\rm F}-q)~.
\eea 
Note that in the final expression for the real part of the extrinsic transverse function in Eq.~\eqref{dopedtrans}, all the terms involving the $\Theta(x)$ terms will cancel each other. 
\end{widetext}

\section{Kramers-Kronig relations and the current-current response function}

To validate the correctness of our current-current response function, we check that they satisfy the Kramers-Kronig relations. 
A response function $f(q,\omega)$, which is analytic in the upper half complex plane and which vanishes in the limit of complex $\omega \to \infty$, satisfies the standard Kramers-Kronig (KK) relation given by, 
\be\label{kk0}
\text{Re}~f(q,\omega)=\frac{1}{\pi}\mathcal{P}\int_{-\infty}^\infty d\xi\frac{\text{Im}~f(q,\xi)}{\xi-\omega}~.
\ee
However in general $f(\omega)$ is of the form $\mathcal{A}(q,\omega)\omega^{n-1}$, with $n$ being a positive integer and 
$\mathcal{A}(q,\omega \to \infty)$ is finite along with $\mathcal{A}(q,\omega \to 0)$. Since the response function $f(q,\omega)$ diverges as $\omega^{n-1}$ in the $\omega \to \infty$ limit, the function $f(q,\omega)/\omega^n$ is used in the Cauchy relation to construct the genralized Kramers-Kronig relation of order $n$ (KK$_n$) with one subtraction \cite{bjorken1965relativistic}.  {The generalized KK$_n$ can be obtained as follows: 
\be \label{E2_1}
\frac{f(\omega)}{\omega^n} = \lim_{\epsilon \to 0} \frac{f(\omega + i \epsilon)}{\omega^n} = \frac{1}{2 \pi i} \oint \frac{f(\xi)}{\xi^n (\xi - \omega)}~.
\ee
Here the contour in the integral is the standard contour in the upper half plane -- parallel to the real axis (just above it) and closed around $\xi \to \infty$ in a semicircle. Since $f(\xi)/\xi^n$ vanishes for $\xi \to \infty$, the r.h.s. of Eq. \eqref{E2_1} has three contributions: 
1) half of the residue of $f(\xi)/\xi^n$ at the point $\xi \to \omega$ and 2) the prinicple value of the integral along the real line and 3) the contribution from the $n$'th order pole at $\xi \to 0$. Combining these three,  we obtain the generalized Kramers Kronig relation to be, 
\be\label{kk01}
\frac{{f(\omega)}}{\omega^n}=\frac{\mathcal{A}(q,0)}{\omega}+\frac{1}{i\pi}\mathcal{P}\int_{-\infty}^\infty d\xi\frac{f(\xi)}{\xi^n(\xi-\omega)}~.
\ee  
The real part of Eq.~\eqref{kk01} is given by 
\be\label{kk1}
\frac{\text{Re}~{f(\omega)}}{\omega^n}=\frac{\text{Re}~\mathcal{A}(q,0)}{\omega}+\frac{1}{\pi}\mathcal{P}\int_{-\infty}^\infty d\xi\frac{\text{Im}~f(\xi)}{\xi^n(\xi-\omega)}, 
\ee  
where $\mathcal{A}(q,0)=\lim_{\xi\to0}[{f(q,\xi)}/{\xi^{n-1}}]$ is the residue of the function $f(z)/z^{n-1}$ at $z=0$. In case the response function is a sum of parts, each part of which has a different power law dependence as $\omega \to \infty$, the different terms have to be considered separately using different KK$_n$ relations.}

{Note that there is an alternate way to construct the generalized Cauchy relations. For $f(\omega)$ does not diverge more than $\omega^{n-1}$ as $\omega\to\infty$, 
instead of using the  the division by $\omega^n$ in $f(\omega)$ to cure the divergence at infinity, we can also use the product $f(\omega)\prod_{m=1}^n\frac{1}{\omega-\omega_m}$ to cure the divergence. 
In this case  an alternate generalized Kramers Kronig relation (of order $n$) can be obtained and it is given by \cite{2017arXiv171101031Z}
\bea\label{eqKK_alt}
f(\omega)\prod_{m=1}^n\frac{1}{\omega-\omega_m}&=&\sum_{l=1}^n\frac{f(\omega_l)}{\omega-\omega_l}\prod_{m=1,m\neq l}^n\frac{1}{\omega_l-\omega_m} \nn \\ 
&+&\frac{1}{i\pi}\mathcal{P}\int_{-\infty}^{\infty}d\xi\frac{f(\xi)}{g(\xi)}+C_\infty
\eea
Where $g(\xi)=(\xi-\omega)\prod_{m=1}^n(\xi-\omega_m)$ and the contribution from the semicircle at infinity vanishes, i.e. $C_\infty \to 0$. Here the choice of $\omega_m$ is arbitrary, and Eq.~\eqref{eqKK_alt} is independent of 
of the choice of $\omega_m$.} 

{It turns out that for a particular choice of $\omega_m$ Eq.~\eqref{eqKK_alt} reduces to Eq.~\eqref{kk1}, which in general is simpler to use. 
For even $n$, we choose $\omega_m =\epsilon\times\{-n/2, -(n-1)/2,...(n-1)/2, n/2 \} $, and for odd $n$ we choose $\omega_m =\epsilon\times\{ -(n-1)/2,..-1,0,1..(n-1)/2\}$. 
Using this choice in Eq.~\eqref{eqKK_alt}, and then taking the limiting case of $\epsilon \to 0$ it is easy to see that the l.h.s. and the second term on the r.h.s. of Eq.~\eqref{eqKK_alt} reduce to corresponding l.h.s. and the second term 
on the r.h.s. of Eq.~\eqref{kk1}. The first term on the r.h.s. of Eq. \eqref{eqKK_alt} reduces to,
\bea \label{term1_1}
\lim_{\epsilon\to0}\sum_{p=0}^{k-1}\frac{(-1)^p(k-p)^{2k}}{p!(2k-p)!}\times\Bigg[\frac{f[\epsilon(k-p)]}{[\epsilon(k-p)]^{n-1}[\omega-\epsilon(k-p)]}\nn\\+\frac{f[-\epsilon(k-p)]}{[-\epsilon(k-p)]^{n-1}[\omega+\epsilon(k-p)]}\Bigg]=\frac{\mathcal{A}(q,0)}{\omega}\nn\\
\eea
where $k=n/2$ and $(n-1)/2$ for even and odd $n$ respectively. In the limiting case of $\epsilon \to 0$, we have for all $m$, 
\be \label{term1_2}
\lim_{\epsilon \to 0}\frac{f(m \epsilon)}{(m \epsilon)^{n-1}(\omega-m \epsilon)} = \frac{{\cal A}(0)}{\omega}.  
\ee
In addition the first term in Eq.~\eqref{term1_2} is simply given by  
\be \label{term1_3}
\sum_{p=0}^{k-1}\frac{(-1)^p(k-p)^{2k}}{p!(2k-p)!}=\frac{1}{2}~. 
\ee
Substituting Eqs.~\eqref{term1_2}-\eqref{term1_3}, in Eq.~\eqref{term1_1} shows that the first term in the r.h.s. of Eq.~\eqref{eqKK_alt} is identical to the first term on the r.h.s. in Eq.~\eqref{kk01}.} 


Let us now consider the intrinsic response first. The imaginary part of the density-density response function is constant at $\omega\to\infty$. So it satisfies KK$_1$ as shown  explicitly in Ref.~[\onlinecite{xiao_arxiv_2014}].
For the longitudinal current-current response function the imaginary part diverges as $\omega^2$. So it satisfies KK$_3$, or more explicitly, 
\bea\label{kk3}
\text{Re}~\Pis^{(0u)}(q,\omega)&=&\omega^2\lim_{\xi\to0}\frac{\text{Re}~\Pis^{(0u)}(q,\xi)}{\xi^2} \\ 
&+&\frac{\omega^3}{\pi}\mathcal{P}\int_{-\infty}^\infty d\xi\frac{\text{Im}~\Pis^{(0u)}(q,\xi)}{\xi^3(\xi-\omega)}~. \nn
\eea 
The first term on the right hand side of Eq.~\eqref{kk3}, can be evaluated by using Eq.~\eqref{reden2spin} and \eqref{chi_roro}, and it is given by, 
\be\label{kk4}
\omega^2\lim_{\xi\to0}\frac{\text{Re}~\Pis^{(0u)}(q,\xi)}{\xi^2}=-\frac{\omega^2}{24\pi^2\hbar v_{\rm F}^3}\log \frac{4q_{\rm max}^2}{q^2}~.
\ee
To evaluate the second term in Eq.~\eqref{kk3} we use Eq.~\eqref{imagden2spin} and Eq.~ \eqref{eq:Imd} to obtain, 
\be\label{kk5}
\frac{\omega^3}{\pi}\mathcal{P}\int_{-\infty}^\infty d\xi\frac{\text{Im}~\Pis^{(0u)}(q,\xi)}{\xi^3(\xi-\omega)}=-\frac{\omega^2}{24\pi^2\hbar v_{\rm F}^3}\log\left| \frac{v_{\rm F}^2q^2}{v_{\rm F}^2q^2-\omega^2}\right|.
\ee
Combining Eqs.~\eqref{kk4}-\eqref{kk5} reproduces the second term in  Eq.~\eqref{reden2spin}. Note that the first term in Eq.~\eqref{reden2spin}, which arises from the anomalous commutator, is real and independent of $\omega$. Thus it trivially  satisfies KK$_1$ with a vanishing imaginary part.

The imaginary part of the intrinsic transverse current-current response function can be split into two parts. First part diverges as $\omega^2$ while the other one is finite for $\omega\to\infty$. Thus the first part obeys KK$_3$ and the second part follows KK$_1$. Following the procedure outlined for the longitudinal case, we have explicitly checked analytically that  the real part of the intrinsic transverse current current response function can be obtained from the corresponding imaginary parts. 

The extrinsic part of both the transverse and the longitudinal response function has no divergence problem and both vanish in the $\omega\to\infty$ limit. Thus the extrinsic part satisfies Eq.~\eqref{kk1}, or alternately the modified version of the 
standard Kramers-Kronig relation given by,
\be\label{kk6}
\text{Re}~\Pis^{(0d)}(q,\omega)= \frac{2}{\pi}\int_0^\infty d\xi \frac{\xi\text{Im}~\Pis^{(0d)}(q,\xi)}{\xi^2-\omega^2}~.
\ee
We have checked numerically that the real part of the response function obtained using Eq.~\eqref{kk6}, are identical to the analytical results for the real part of the longitudinal and transverse response function in Eq.~\eqref{eq:16} and Eq.~\eqref{dopedtrans}, respectively.

\section{QED and charge renormalization}
\label{QED}
The QED Lagrangian with arbitrary scaling factors for the vector field ($A^\mu \to \sqrt{Z_3} A^\mu_r$) and the spinor wavefunction ($\psi \to \sqrt{Z_2} \psi_r)$, is given by 
\be \label{Lag}
{\cal L}=-\frac{1}{4} Z_3 F^r_{\mu \nu}F^r_{\mu \nu} + i Z_2 \bar{\psi_r} \gamma^\mu \partial_\mu \psi_r + \sqrt{Z_3} Z_2 e \bar{\psi_r} \gamma_\mu A^\mu_r \psi_r~,
\ee
where $e$ denotes the bare charge. Here $Z_3$, $Z_2$  are determined by renormalization to cancel the anomalous diverging terms in the loop integrals. 
Here the last term can be expressed in terms of renormalized charge by defining $e_r Z_1 = e \sqrt{Z_3} Z_2$. 
Now the Ward-Takahashi identity implies that $Z_1 ={Z_2}$. This also allows the last two terms of Eq.~\eqref{Lag} to be expressed as a covariant derivative: $i Z_2 \bar{\psi_r} \gamma^\mu \partial_\mu \psi_r + \sqrt{Z_3} Z_2  e \bar{\psi_r} \gamma_\mu A^\mu_r \psi_r~ = i Z_2 \bar{\psi_r} \gamma^\mu (\partial_\mu - e_r A_\mu^r) \psi_r$. 

The single loop polarization diagram or the self energy correction to the photon propagator [intrinsic current-current correlator in Eq.~\eqref{reden2spin}] has an anomalous diverging term $-(q_{\rm max}^2 e_r^2 v_{\rm F}/({6\pi^2\hbar})$. This  implies that the appropriate counter terms to cancel it, which renormalizes the vector field is given by 
\be
\sqrt{Z_3} = \left(1- \frac{e_r^2 v_F q_{\rm max}^2}{6\pi^2\hbar}\right)^{1/2}~.
\ee
Consequently the bare and the renormalized charges are related by 
\be
e = e_r \times \left(1- \frac{e_r^2 v_F q_{\rm max}^2}{6\pi^2\hbar}\right)^{-1/2}~.
\ee
The renormalized charge $e_r$ is the scale dependent effective screened charge which is observed in experiments.
\bibliography{CC_corr_3D_refs}
\end{document}